\documentclass[a4paper,11pt]{article}

\begin{document}
\def\boxit#1{\vcenter{\hrule\hbox{\vrule\kern8pt
      \vbox{\kern8pt#1\kern8pt}\kern8pt\vrule}\hrule}}
\def\Boxed#1{\boxit{\hbox{$\displaystyle{#1}$}}} 
\def\sqr#1#2{{\vcenter{\vbox{\hrule height.#2pt
        \hbox{\vrule width.#2pt height#1pt \kern#1pt
          \vrule width.#2pt}
        \hrule height.#2pt}}}}
\def\square{\mathchoice\sqr34\sqr34\sqr{2.1}3\sqr{1.5}3}
\def\Square{\mathchoice\sqr67\sqr67\sqr{5.1}3\sqr{1.5}3}
\def\AM{{\it Ann. Math.}}
\def\AP{Ann. Phys.}
\def\CQG{{\it Class. Quantum Grav.}}
\def\GRG{{\it Gen. Rel. Grav.}}
\def\JMP{{\it J. Math. Phys.}}
\def\JP{J. Phys.}
\def\JSIRAN{{\it J. Sci. I. R. Iran}}
\def\NC{{\it Nuovo Cim.}}
\def\NP{{\it Nucl. Phys.}}
\def\PL{{\it Phys. Lett.}}
\def\PR{{\it Phys. Rev.}}
\def\PRL{Phys. Rev. Lett.}
\def\PRp{{\it Phys. Rep.}}
\def\RMP{{\it Rev. Mod. Phys.}}
\title{On Higher Order Gravities, Their Analogy to GR, and
       Dimensional Dependent Version of Duff's Trace Anomaly
       Relation}

\author{{\small Mehrdad Farhoudi}\footnote{m-farhoudi@sbu.ac.ir}\\
        {\small Department of Physics, Shahid Beheshti University,}\\
        {\small Evin, Tehran 1983963113, Iran}}

\date{\small October 31, 2005}

\maketitle

\begin{abstract}
An almost brief, though lengthy, review introduction about the
long history of higher order gravities and their applications, as
employed in the literature, is provided. We review the analogous
procedure between higher order gravities and GR, as described in
our previous works, in order to highlight its important
achievements. Amongst which are presentation of an easy
classification of higher order Lagrangians and its employment as a
\emph{criteria} in order to distinguish correct metric theories of
gravity. For example, it does~not permit the inclusion of only one
of the second order Lagrangians in \emph{isolation}. But, it does
allow the inclusion of the cosmological term. We also discuss on
the compatibility of our procedure and the Mach idea. We derive a
dimensional dependent version of Duff's trace anomaly relation,
which in \emph{four}-dimension is the same as the usual Duff
relation. The Lanczos Lagrangian satisfies this new constraint in
\emph{any} dimension. The square of the Weyl tensor identically
satisfies it independent of dimension, however, this Lagrangian
satisfies the previous relation only in three and four dimensions.
\end{abstract}

\bigskip

\section{Higher Order Gravity Review}
\indent

In spite of the success of Einstein's gravitational theory, it can
only be considered as a step towards a much more complete and
comprehensive structure due to a number of seeming weaknesses.
Even its own past successes have also raised deep and difficult
problems, e.g. it has not yet been brought into a unified theory
of nature~\cite{ishishb}. The failure to establish a proper
quantum theory of gravity presently poses basic questions
concerning the scope and general formalism of such a theory
itself, and induces a search for alternative theories of
gravitation~\cite{negogoefran}. Interest in this subject stems
mostly from making a systematic investigation into the foundations
of general relativity amongst which are fundamental changes in
dynamics and dimension of space--time.

The principle of general invariance imposes that the action
integral for gravity must be an invariant quantity, indeed, the
profound implications of this state of affairs are due to
Hilbert~\cite{hil}\ who was first to employ the necessity of the
general invariance of all physical laws to the variational
principle. Then, the use of the action principle and the principle
of general invariance allow an immediate connection between
symmetry principles and conservation laws to be established as
inner identities. Yet, change of dynamics can mainly be achieved
by altering Lagrangian of a system, as in general relativity~(GR)
the physical requirements which should be satisfied by any
reasonable Lagrangian formulation of gravitational interactions
are still an open subject for discussion.

Among scalar Lagrangians, field equations based on a Lagrangian
quadratic in the curvature tensor have had a long history in the
theory of gravitation. These theories occur for various reasons in
different areas of theoretical physics. Though, they have taken
much more attraction in the last two decades when they have been
proposed mainly to solve some problems in the quantization of
gravity. But, the first idea dates back to the early days of GR in
the work by Weyl~\cite{weyl}\ and Eddington~\cite{edd}\ in an
attempt to unify gravity with electromagnetism, however this
approach was unfruitful~\cite{stelb,malu}.

Then, Bach~\cite{bac}\ and Lanczos~\cite{lanalanaa}\ considered
actions that are made up \emph{only}\ by the quadratic terms in
the curvature in four dimensions as scale-invariance, due to
Weyl's idea of the principle of gauge-invariance under conformal
transformations for when the conformal factor is a constant, e.g.
they studied the dimensionless action $\int R^2\,\sqrt{-g}\,
d^4\!x$. However, this idea has also been strongly criticized as a
nonviable theory~\cite{buchdcbick}. The two main objections
against these Lagrangians are as the metric based on them does~not
satisfy the flat space limit at asymptotically large distances,
and disagreement with observations follows when the matter is
incorporated. A Lagrangian proportional to $R^{1+ \delta}$ has
also been investigated~\cite{cliftonba}, and it has been shown
that the observational constraints leads to the overall bound
$0\leq \delta < 7.2\times 10^{-19}$. Exact cosmological solutions
of this type of scale--invariant gravity theories have been
considered in~\cite{barrowcl}.

Since these early suggestions, work in this area was~not very
active, although some results were obtained\footnote{See, for
 example, Ref.~\cite{buch70}.}\
before it flourished in the seventies, i.e. after suggestions made
from considering the quantization of GR in the sixties.

Perhaps a legitimate mathematical motivation to examine
gravitational theories built on non-linear Lagrangians has been
the phenomenological character of Einstein's theory which leaves
room for such amendments, i.e. the dependence of the Einstein
tensor/Lagrangian on the derivatives of the metric and the
dimension. Actually, the Einstein Lagrangian is~not the most
general second order Lagrangian allowed by the principle of
general invariance, and indeed, through this principle the latter
generalization can be performed up to \emph{any}\ order, and a
general scalar Lagrangian is a higher derivative Lagrangian.

However, Einstein's gravitational theory has proven to be a
successful theory in ``real'' physical phenomena, e.g. in the
``weak'' gravitational field, and its main difficulties become
manifest only when the curvature of space--time is~not negligible.
Actually, the curvature is noticeable on very small scales, and
this is particularly relevant to the very early universe. In
another words, at really small distances, of the order of Planck's
length, an Euclidean topological structure is quite unlikely. At
such distances, even the fluctuations of quantum gravitation will
be extremely violent and probably produce an ever changing,
dynamic topology~\cite{wheb}. Thus, it perhaps allows Lagrangians
extended beyond the Einstein--Hilbert Lagrangian to be considered
as alternative theories.

On the other hand, the unity of physics during its development
must be maintained by the correspondence principle. That is, in
every ``new'' physical theory the ``previous'' one is contained as
a ``limiting'' case. As mentioned before, gravitational theories
based on a Lagrangian which is \emph{only}\ purely quadratic in
the curvature tensor have been strongly
criticized~\cite{buchdcbick}\ as nonviable theories. However, a
gravitational theory has to not only correctly reflect the
dynamical behavior of the whole universe, but also be valid for
stellar evolution. Therefore, one should demand that GR must be
maintained as a ``limiting'' case of non-linear gravitational
theories.

Evidently, the unification programs in particle physics seem to
demand energies of the order of cosmological energies, $E\geq
10^{15}\ GeV$, in order to be verified, and thus the very early
universe may provide the only place to test these ideas. Hence,
the standard prejudices about the strength of quantum effects on
gravitational interaction imply that they should~not be
appreciable unless the distances involved in the problem are very
small, of the order of Planck's length. So, the questions of
quantum gravity are naturally connected with the very early
universe cosmology~\cite{ishb}. That is, these effects should be
completely negligible in the ``limiting'' case of non-linear
gravitational Lagrangians, i.e. in the low energy regime when
$E\ll E_p\approx 10^{19}\ GeV$.

It is believed that higher order Lagrangians play an essential
r\^ole for supergravity approach~\cite{cfgpp}. Nowadays, it is
also well known\footnote{See, for example, Refs.~\cite{bidabos}\
and references
 therein. }\
that Einstein's gravity when treated as a fundamental quantum
gravity leads to a non-renormalizable theory, although, these
difficulties become manifest only when the curvature of
space--time is~not negligible. In order to permit renormalization
of the divergences, quantum gravity in curved space--time has
indicated that the Einstein--Hilbert action should be enlarged by
the inclusion of higher order curvature counter
terms~\cite{utdepese}. In fact, it has been shown$^2$ that the
Lagrangian\footnote{$\alpha$,
 $\beta$ and
 $\kappa^2\equiv \frac{16\pi G}{c^4}$ are constants, and where
 field equations are shown as
 $G^{(\rm gravitation)}_{\alpha\beta}={1\over 2}\kappa^2\,
 T_{\alpha\beta}$. }\
\begin{equation}\label{eplii}
L=\frac{1}{\kappa^2}\bigl(R+\alpha R^2+\beta
    R_{\mu\nu}R^{\mu\nu}\bigr)\ ,
\end{equation}
which has the required Newtonian limit and, by the Gauss--Bonnet
theorem (relation)~\cite{cheachebkonospiv}, is the most general
quadratic Lagrangian in (and up to) \emph{four} dimensions, solves
the renormalization problem and is multiplicatively
renormalizable~\cite{stela}\ and asymptotically free~\cite{frtsa}.
However, it is~not unitary within usual perturbation
theory~\cite{stelb}. The analysis of quantum theory
revealed~\cite{stela} that the particle spectrum of the
Lagrangian~(\ref{eplii}), in general, contains, besides the
massless spin-two particles (i.e. gravitons), a further massive
spin-two particle (i.e. ghost) and a massive scalar (spin-zero)
particle, with a total of eight degrees of freedom. In the limit
$\beta\to 0$, the spin-two ghost disappears, however, the
divergence problems in the quantum theory reappear~\cite{stelb}.

The inclusion of higher order Lagrangian terms have also appeared
in the effect of string theory on classical gravitational physics
by means of a low energy effective action which expresses gravity
at the classical level~\cite{chswmtgswltkp,dese}. This effective
action in general gives rise to fourth order field equations (and
brings in ghosts), and in particular cases, i.e. in the form of
dimensionally continued Gauss--Bonnet densities, it is exactly the
Lovelock terms\footnote{The Lovelock Lagrangian, in a $D$
dimensional space--time, is~\cite{lovedc,brig}
 \begin{equation}\label{love}
 {\cal L}=\frac{1}{\kappa^2}\sum_{0<n<\frac{D}{2}}\,\frac{1}{2^n}\,
 c_n\,\delta^{\alpha_1\ldots\alpha_{2n}}_{\beta_1\ldots\beta_{2n}}\,
 R_{\alpha_1\alpha_2}{}^{\beta_1\beta_2}\cdots R_{\alpha_{2n-1}\,
 \alpha_{2n}}{}^{\beta_{2n-1}\,\beta_{2n}}
 \equiv\sum_{0<n<\frac{D}{2}}\, c_n\, L^{(n)}\ ,
 \end{equation}
 where we set $c_1\equiv 1$ and the other $c_n$ constants to be of
 the order of Planck's length to the power $2(n-1)$, for the
 dimension of $\cal L$ to be the same as $L^{(1)}\equiv L_{_{E-H}}=R$.
 $\delta^{\alpha_1\ldots\alpha_p}_{\beta_1\ldots\beta_p}$ is the
 generalized Kronecker delta symbol, which is identically zero if
 $p>D$ and the maximum value of $n$ is related to the dimension of
 space--time by
 \begin{equation}\label{nlim}
 n_{_{\textrm{max}}}\!= \cases{
                          \frac{D}{2}-1 & \textrm{even $D$}\cr
                          \cr
                          \frac{D-1}{2} & \textrm{odd $D$.}\cr}
 \end{equation}
 Hence, $\cal L$ reduces to the Einstein--Hilbert Lagrangian in
 four dimensions.\newline
 Implicitly , we follow the sign conventions of Wald~\cite{wald}. }\
(and consequently no ghosts arise)~\cite{zwizum}. However, Duff
{\it et al}\ \cite{dnpb}\ claimed that the quadratic contribution
to the low energy effective Lagrangians for the closed bosonic
string is of the form $R^2-4\, R_{\mu\nu}R^{\mu\nu}+2\,
R_{\alpha\beta\mu\nu}\, R^{\alpha\beta\mu\nu}$, for the Type II is
of the form $R^2-4\, R_{\mu\nu}R^{\mu\nu}$ and only for the
$E_8\times E_8$ heterotic string do form the Gauss--Bonnet
combination. The latter term appears naturally in the
next-to-leading order term of the heterotic string effective
action and also plays an essential role in the Chern--Simons
gravitational theories~\cite{chammuh,allfrr}.

Incidentally, recent argument in~\cite{nood}\ shows that $1\over
R$ modification term also follows from certain compactification
schemes of string/M--theory.

Though, higher derivative terms are often added as a correction to
the ordinary, lower derivative, theory of gravity, but these terms
do~not only mean that they will perturb the original theory.
Actually, their presence, as unconstrained terms even with small
coefficients, make the new theory completely different from the
original one~\cite{simo}.

Besides, there has been much attraction in considering gravity in
higher dimensional space--time. In this context, one may also use
a consistent theory of gravity with a more general action, e.g.
the Einstein--Hilbert action plus higher order terms. Indeed, the
above ghost-free property, and the fact that the Lovelock
Lagrangian is the most general second order Lagrangian which, the
same as the Einstein Lagrangian, yields the field equations as
{\it second} order equations, have stimulated interests in the
Lovelock gravity and its applications. A considerable amount of
work has been done in this area, especially in the eighties, and
perhaps, the greatest number of applications are performed in
cosmology~\cite{mdmmmpcf},\footnote{For a more recent work on this
 issue see, e.g. Ref.~\cite{cdct}.}\
amongst which are approaches
to inflationary scenarios\rlap.\footnote{See, for example,
 Refs.~\cite{berm}--\cite{goms} and references therein. }\
Recent attraction engages it with the brane cosmology and the
cosmic acceleration~\cite{dehamengwangabf,dehb}.

Actually, the observation of high red-shift
supernova~\cite{rptetal}\ and the measurement of angular
fluctuations of cosmic microwave background
fluctuations~\cite{lnhsetal}\ have separately, though the former
directly and the latter indirectly, established that the universe
expands with acceleration, instead of deceleration, at the present
epoch. Henceforth, as an alternative approach to the vacuum energy
(the cosmological constant) and/or additional scalar field
(quintessence), modifications of gravity itself with different
terms of higher order gravity have recently come into
consideration~\cite{dehb} , \cite{cdttccct}--\cite{noodc}, for,
e.g., the $1\over R$ modified term grows when curvature decreases
and it may produce the cosmic acceleration. However,
Ref.~\cite{sousawodard04} claims that they have found a linearly
growing force which is~not phenomenologically acceptable. Besides,
the $1\over R$ term modification has been shown to lead to
instabilities~\cite{chiba03}--\cite{dolgovkawasaki2003}. Though,
Ref.~\cite{nooda}\ claims that further modification of this
modified gravity by $R^2$ or other higher order terms may resolve
the instabilities, or perhaps make them
avoidable~\cite{noodb,noodc}.

The charges associated to the diffeomorphism symmetries of
Lovelock gravity in any odd dimensional space--time has been
investigated in~\cite{allfrr}. The boundary term~\cite{maba},
Lovelock--Cartan gravity~\cite{marza}\ and Palatini's
device~\cite{palastepe}\ for non-linear Lagrangians have been
considered as well. Though, Buchdahl~\cite{buchde}\ has claimed
the mutual in-equivalence of H-variation and P-variation, and has
illustrated this for quadratic Lagrangians as an example. However,
it has been shown~\cite{bshs} that the latter Lagrangian still
satisfies Birkhoff's theorem. For recent work on this issue, see
e.g. Refs.~\cite{mengwangsotiriou}. Besides, it has been
claimed~\cite{volikmengwangcapoetal} that the Palatini formalism
of the $1\over R$ modified gravity has no such mentioned
instabilities, but still provides the current observations.
Actually, Ref.~\cite{olmok04} claims that the Palatini formalism
can provide a mechanism to explain the cosmic acceleration without
the necessity of dark energy sources. However, new instabilities
from quantum effects may appear~\cite{flanaganmengwang}. But,
Ref.~\cite{noodc} shows these new instabilities may be suppressed
by quantum effects of conformal fields, and
Ref.~\cite{vollick2004} argues that the conclusion for the new
instabilities is false, for mathematical equivalence, and not
physical equivalence, has been used.

Non-linear Lagrangians are also proposed in the context of
relativistic cosmology in order to avoid the appearance of some of
the singularities typically encountered for solutions of
GR\rlap.\footnote{See, for
 example, Refs.~\cite{kermadhost}. }\
The general behavior of cosmology, the existence and stability of
special solutions, like de Sitter, have been considered for
theories with $f(R)$ Lagrangian in, e.g., Refs.~\cite{baotmuss}.
For recent work on this issue see Ref.~\cite{faraoni05} and
references therein.

Because of the higher order and greater degree of the
non-linearity of the field equations, it is very difficult to find
out non-trivial exact analytical solutions of Einstein's equation
with higher order terms, and hence, extract physical predictions
from them, especially that the present technology hardly may
provide empirical checks on these predictions. Though, the
Newtonian limit of fourth~\cite{stelb,tey}\ and higher~\cite{qus}\
order gravities are described by a Newtonian plus Yukawa term
potential. Recently, the post--Newtonain parameters limit of such
theories, especially $f(R)$ Lagrangian with corrections that grow
at low curvatures, has been analyzed both in the metric and in the
Palatini
formalisms~\cite{chiba03,sousawodard04,dmwdbco3afrts,cembra05}
with contrasting results for being compatible with solar system
observations. As, e.g., Ref.~\cite{cembra05} claims that the
analyzes performed in some works are~not founded since they are
obtained at the low energy approximation, which is~not the case
inside the entire solar system. For a more recent work on this
issue see Ref.~\cite{capoztroi05}.

The Cauchy problem for the fourth order field equations deriving
from $R+\alpha\, R^2$ is essentially reducible to a second order
one~\cite{teto} and its energy is a positive definite
quantity~\cite{strmaluab}. The positive energy theorem has also
been shown to hold for a larger class of non-linear Lagrangians
for which the Einstein frame can be defined around flat
space~\cite{magsok}.

In general, the reduction of non-linear purely metric Lagrangian
theories of gravity to a second order form, as equivalent theories
to GR plus additional matter fields, so-called ``dynamical
universality'' of Einstein's gravity, under conformal/Legendre
like transformations, has been discussed in the literature, see,
e.g., Refs.~\cite{mffabsirz}\ and references therein. Although, it
has raised a question on the real physical metric~\cite{bran},
where, in reply~\cite{ffmabso}, comments are in having the
dynamical equivalence of non-singular conformal transformations
where then, the systems are isomorphic, and not having necessarily
the physical equivalence\rlap.\footnote{Also, see
 Ref.~\cite{pasa}.}\
Also, the dynamical equivalence of arbitrary high order extended
theories of gravity with scalar--tensor gravity, which claims to
be conformally equivalent to GR plus scalar fields, has been
considered~\cite{tey,bacomaewandhiow,goss}. More discussion about
physical equivalence of non-linear gravity theories, based on the
physical features of energy, has been presented in
Ref.~\cite{magsok}.

Some static black hole solutions of the Gauss--Bonnet gravity have
been obtained in~\cite{bdcsnoocnocacn}. And equations governing
relativistic fluid dynamics for quadratic theories and $f(R)$
theories of gravity have also been worked out~\cite{mtbnovrett}.

Finally, considering the corresponding Euler--Lagrange expressions
of Lagrangians containing the \emph{derivatives} of the curvature
scalar are firstly due to Buchdahl~\cite{buchd}. In general,
$f\bigl(\sum\limits^p_{k=0} \Square\,^k R\bigr)$ Lagrangian, where
$p$ is a positive integer and among which is the well known sixth
order gravity~\cite{berm,goms,goss,buchd} and its
generalization~\cite{aetal}, has been considered in the literature
as well. The corresponding Euler--Lagrange expression for the
special case of $f(R)$ has been given in~\cite{stepd}, and the
power series example, i.e. $f\bigl(R\bigr)=\sum\limits^p_{k=0}
a_k\, R^k$, has also been used in~\cite{roxcol}.

Yet, very vast amount of work has been done in the higher order
gravities and the references given in this compact survey {\it are
not} obviously a complete bibliography on this issue. Actually,
this section is almost brief, though lengthy, review introduction
about the long history of higher order gravities and their
applications. In the next section, we will introduce what we have
achieved on this subject.

\section{Introduction}
\indent

Getting interested in higher order gravities, we have noticed that
the mathematical form of the Einstein tensor, i.e. the splitting
feature of it into the Ricci tensor and the curvature scalar with
the {\it trace} relation between them is also a common remarkable
properties of {\it each} homogeneous term in the Lovelock
tensor\rlap,\protect\footnote{That is $G^{(n)}_{\alpha\beta}$,
 where the Lovelock tensor, as dimensionally reduction
 Euler--Lagrange terms in a $D$ dimensional space--time, is~\cite{lovedc,brig}
 \begin{equation}\label{lovet}
 {\cal G}_{\alpha\beta}\!=-\!\!\!\!\sum_{0<n<\frac{D}{2}}
 \!\!\frac{1}{2^{n+1}}\, c_n\, g_{\alpha\mu}\,
 \delta^{\mu\alpha_1\ldots\alpha_{2n}}_{\beta\beta_1\ldots\beta_{2n}}
 \, R_{\alpha_1\alpha_2}{}^{\beta_1\beta_2}\cdots
 R_{\alpha_{2n-1}\,\alpha_{2n}}{}^{\beta_{2n-1}\,\beta_{2n}}\equiv
 \!\!\!\!\sum_{0<n<\frac{D}{2}}\!\! c_n\, G^{(n)}_{\alpha\beta}
 \end{equation}
 where the cosmological term has been neglected and
 $G^{(1)}_{\alpha\beta}\equiv G_{\alpha\beta}$, i.e. the Einstein tensor. }\
but not true for the whole Lovelock tensor. In our previous
work~\cite{farb}, we demanded an analogy that this common feature
should also be valid for the Lovelock tensor itself, i.e. ${\cal
G}_{\alpha\beta}=\Re_{\alpha\beta}-\frac{1}{2}
g_{\alpha\beta}\,\Re$\ with ${\sl Trace}\, \Re_{\alpha\beta}=\Re$,
or generally, for any {\it inhomogeneous} Euler--Lagrange
expression constructed {\it linearly} in terms of homogeneous
terms. Indeed, we did this via defining a generalized trace
operator, that we denoted by ${\sl
Trace}$\rlap.\protect\footnote{That
 is, for a general $\bigl({N\atop M}\bigr)$ tensor which is a
 homogeneous function of degree $h$ with respect to the metric and
 its derivatives, we defined
 \begin{equation}\label{Tracedu}
 \textrm{Trace}\, {}^{[h]}A^{\alpha_1\ldots\alpha_N}
    {}_{\beta_1\ldots\beta_M}
    :=\!\cases{
       \frac{1}{h-\frac{N}{2}+\frac{M}{2}}\, \textrm{trace}\,
       {}^{[h]}A^{\alpha_1\ldots\alpha_N}{}_{\beta_1\ldots\beta_M} &
       \textrm{when $h-\frac{N}{2}+\frac{M}{2}\not=0$}\cr
       \textrm{trace}\, {}^{[h]}A^{\alpha_1\ldots\alpha_N}
       {}_{\beta_1\ldots\beta_M} &
       \textrm{when $h-\frac{N}{2}+\frac{M}{2}=0$.}\cr}
 \end{equation}
 Hence, for example, when $h\neq 0$ and $h'\not=0$, one gets
 \begin{equation}\label{Tracexampl}
 \textrm{Trace}\,\bigl({}^{[h']}C\, {}^{[h]}A_{\mu\nu}\bigr)=\cases{
     \frac{h+1}{h'+h+1}\, {}^{[h']}C\,\textrm{Trace}\,
     {}^{[h]}A_{\mu\nu} &
     \textrm{for $h\not=-1$}\cr
     \frac{1}{h'}\, {}^{[h']}C\,\textrm{Trace}\,
     {}^{[h]}A_{\mu\nu} &
     \textrm{for $h=-1$.}\cr}
 \end{equation}
 As mentioned, the homogeneity is taken with respect to the metric and
 its derivatives, without loss of generality, with the {\it homogeneity
 degree number} ({\bf HDN}) conventions of ${}^{[+1]}g^{\mu\nu}$ and
 ${}^{[+1]}g^{\mu\nu}{}_{,\alpha}$. Hence, one can relate the
 orders $n$ in any Lagrangian, as in $L^{(n)}$, that represents its
 HDN. See Ref.~\cite{farb} for details. }\

Then, we took~\cite{farc}\ the above analogy further for the {\it
alternative} form of the Lovelock equation to be the same as the
appearance of the alternative form of the Einstein equation.
Hence, we found that the price for this analogy is to accept the
existence of the trace anomaly of the energy-momentum tensor even
in classical treatments. That is, we have actually stated a
classical view of gravitation which explicitly shows the presence
of an extra (anomalous) trace for the energy-momentum tensor, with
an indication of the constitution of the higher order gravities
towards this trace anomaly, exactly as what has been verified in
the quantum aspects of gravity~\cite{dufc,asgosh}.

As an example, we employed~\cite{farc}\ this analogy to any
generic, the most general, second order Lagrangian
\begin{equation}\label{lii}
L^{(2)}_{\rm generic}={1 \over\kappa^2}\Bigl(a_1 R^2+a_2
R_{\mu\nu}R^{\mu\nu}+a_3 R_{\alpha\beta\mu\nu}\,
       R^{\alpha\beta\mu\nu}\Bigr)\ ,
\end{equation}
in $D\geq 3$ dimension\rlap.\footnote{The $a_1$, $a_2$ and $a_3$
 are arbitrary dimensionless constants, and obviously, in three and
 four dimensions, only two of these three terms are effective. And in
 two dimensions, only one term survives.}\
Its corresponding Euler--Lagrange expression is
{\setlength\arraycolsep{2pt}
\begin{eqnarray}\label{nlii}
&G&\!^{(2)}_{({\rm generic})\alpha\beta}\! \equiv \!{\kappa^2\over
   \sqrt{-g}}\,{\delta\bigl(L^{(2)}_{\rm generic}\sqrt{-g}\bigr)\over
   \delta g^{\alpha\beta}}\cr
&=&\!\!2\biggl[a_1 R\, R_{\alpha\beta}
   -2a_3 R_{\alpha\mu}\, R_{\beta}{}^{\mu}
   +a_3 R_{\alpha\rho\mu\nu}\, R_{\beta}{}^{\rho\mu\nu}
   +\bigl(a_2+2a_3\bigr)R_{\alpha\mu\beta\nu}\,
   R^{\mu\nu}\cr
&&{}\quad -\bigl(a_1+\frac{1}{2}\,a_2+a_3\bigr)\,
   R_{;\,\alpha\beta}
   +\bigl(\frac{1}{2}\,a_2+2\,a_3\bigr)\,
   \Square\, R_{\alpha\beta}\biggr]\cr
&&\!\!-\frac{1}{2}\, g_{\alpha\beta}\biggl[\kappa^2\,
   L^{(2)}_{\rm generic}-\!\bigl(4\,a_1+a_2\bigr)\,
   \Square\, R\biggr]\ ,
\end{eqnarray} }
where $\Square \equiv {}_{;\,\rho}{}^\rho$, and where it can be
written as
\begin{equation}\label{gtgeneric}
G^{(2)}_{({\rm generic})\alpha\beta}\equiv R^{(2)}_{({\rm generic})
\alpha\beta} -{1\over 2}g_{\alpha\beta}\, R^{(2)}_{\rm generic}\ .
\end{equation}
Then, in order that the ``trace'' relation, ${\sl Trace}\,
R^{(2)}_{({\rm generic})\alpha\beta}=R^{(2)}_{\rm generic}$, to be
satisfied, we exactly derived the trace anomaly relation
\begin{equation}\label{cnlii}
3\, a_1+a_2+a_3=0\ .
\end{equation}
which, in the process of re-examining the Weyl anomaly's
applications, was also suggested by Duff~\cite{dufa} in {\it any}
dimension.

Note that, as a \emph{homogeneous} Euler--Lagrange expression has
a \emph{uniform} HDN, then one can work with the usual {\sl trace}
instead of the {\sl Trace} operator, and obtains the same results
if one demands the appropriate ``trace'' relation, i.e. ${1\over
n}{\sl trace}\, R^{(n)}_{({\rm generic})\alpha\beta}=R^{(n)}_{\rm
generic}$. Though, the notion of Trace operator was introduced as
to be able to deal with when one considers the Einstein--Hilbert
Lagrangian \emph{plus} higher order \emph{terms} as a complete
gravitational Lagrangian, i.e. when one works with an
\emph{inhomogeneous} Lagrangian constructed linearly in terms of
homogeneous terms.

One may conclude that the appearance of trace anomaly maybe
interpreted as Lovelock modification of gravity even in classical
treatments. Though, the above consistency condition has also been
derived~\cite{bocr,bopb}\ based on a {\it cohomological} point of
view, using the Wess--Zumino consistency conditions, which was
claimed to be the true reason for the existence of such a
relation. It has also been recently claimed that the AdS/CFT
correspondence, namely the holographic conformal anomaly, maybe
responsible for it, see e.g. Ref.~\cite{noodog}\ and references
therein.

The extra trace of the energy-momentum tensor for generic cases
is~\cite{farc}
\begin{equation}\label{dtng}
T'=-\kappa^{-2}\,D\,\sum_{n\geq 1}\,{n-1\over n}\, c_n\,
   R^{(n)}_{\rm generic}\equiv \sum_{n\geq 1}\, T'_n\ ,
\end{equation}
where $R^{(1)}_{\rm generic}\equiv R$\, (however, $T'_1=0$), and
there is no upper limit for $n$ in generic cases.

For the above example of $n=2$, it is
\begin{equation}\label{dttg}
T'_2=-{\kappa^{-2}\,D\over 2}\, c_2\, R^{(2)}_{\rm generic}\ ,
\end{equation}
where $R^{(2)}_{\rm generic}$, according to equations~(\ref{nlii})
and (\ref{gtgeneric}), is
\begin{equation}\label{rgii}
R^{(2)}_{\rm generic}\equiv \kappa^2\,L^{(2)}_{\rm
generic}-\bigl(4\, a_1+a_2\bigr)\,\Square\, R\ ,
\end{equation}
with constraint~(\ref{cnlii}). The relation~(\ref{dttg}) is
exactly the same as the relevant most general form of the
anomalous trace of the energy-momentum tensor for classically
conformally invariant fields of arbitrary spin and dimension which
has been shown~\cite{ddichr}\ to be
\begin{equation}\label{fanom}
\big\langle T_\rho{}^\rho\big\rangle_{\rm ren}=-{\hbar\, c\over
180(4\pi)^2}\Bigl(a_1 R^2+a_2 R_{\mu\nu}R^{\mu\nu}+a_3
R_{\alpha\beta\mu\nu}\, R^{\alpha\beta\mu\nu}+\gamma\,\Square\,
R\Bigr)\ .
\end{equation}
This obviously shows that
\begin{equation}\label{dodelta}
\gamma=-\bigl(4\, a_1+a_2\bigr)\ ,
\end{equation}
which it completes the trace anomaly relation suggested by
Duff~\cite{dufa}. However, the calculations by most authors
confirmed~\cite{dufa}\ the constraint~(\ref{cnlii}) for the values
of $a_1$, $a_2$ and $a_3$ in all cases, but not always that of
$\gamma$. Apparently, this is due to the fact that the term
$\,\Square\, R$ is a local anomaly~\cite{dufa}. The ambiguities of
the four dimensional anomaly has been recently investigated in
Ref.~\cite{asgosh}.

Actually, in the semi-classical approach of quantum gravity theory
employed to deduce the trace anomalies, the effective action
is~\cite{avra}\ a {\it covariant functional} i.e., invariant under
diffeomorphisms and local gauge transformations. Therefore, the
approximation procedures for calculating the effective action have
to preserve the general covariance at {\it each order}. Hence,
conformal invariance~\cite{frw}\ is also sacrificed to the needs
of general covariance, i.e. the Weyl conformal invariance is~not a
{\it good} symmetry beyond the classical level. This is what we
have also performed in the classical theory of gravitation through
preserving the covariant property of the linear Lagrangian theory
of Einstein's gravity for each order of non-linear theories of
gravitation, by an analogous demand.

Hence, the origin of Duff's suggested relation between the
coefficients of the conformal anomalies may {\it classically} be
interpreted due to the general covariance of Einstein's theory as
applied to the second order of the Lovelock modification of
gravity. Though, it is somehow a naive conjecture, nevertheless,
it gives almost an easy classical procedure to grasp the desired
result, which, (perhaps) besides what have been already given in
the literature~\cite{bocr,bopb,noodog}, may indicate of an
intrinsic property behind it.

In this article, first in the following section, in order to
present a better view of the analogy/approach that we have
employed, we will summarize its important achievements, which
mainly have~not been highlighted before. Then, in Section~4, based
on this analogy, we will derive a \emph{dimensional dependent}
version of Duff's suggested relation for trace anomaly.

\section{Remarks}
\indent

In this section, we will outline what mainly have been gained in
the above mentioned analogous procedure till now, which have~not
been emphasized or stated in our previous works. Aimed to show, as
a main advantage, that this analogous of the Einstein tensor can
be employed as a \emph{criteria} in order to distinguish
\emph{correct/legitimate} metric theories of gravity, which are
either homogeneous functions or \emph{linear}\ combinations of
different homogeneous functions of the metric and its derivatives.

These remarks are as follows:

\begin{itemize}
\item If one allows scalar Lagrangian to be up to the $k^{th}$ order
      jet-prolongation of the metric, with $k\geq 2$, there would be a lot
      of Lagrangian choices that can be considered. For example, higher--loop
      quantum corrections to GR are expected to contain
      terms of the type $R\, \Square\,^k R$ in the Lagrangian~\cite{dese,goss}.
      That is, the gravitational theory with the Lagrangian
      \begin{equation}\label{msixo}
      L={1\over \kappa^2}\Bigl(R+\sum^p_{k=0}\, a_k\, R\, \Square\,^k
      R\Bigr) \ ,
      \end{equation}
      where $p$ is a positive integer and $a_k$'s are constants of the order
      of (Planck's) length to the power $2k+2$. The above Lagrangian is up
      to $(2p+2)^{th}$ order jet-prolongation of the metric which, in general,
      leads to
      the gravitational field equations of order $2p+4$. For $p=1$, the
      Lagrangian is
      \begin{equation}\label{sixo}
      L={1\over \kappa^2}\Bigl(R+a_0\, R^2+a_1\, R\, \Square\, R\Bigr)\ ,
      \end{equation}
      which is known~\cite{goss}\ as sixth order gravity in the
      literature, i.e. it leads to sixth order field equations, and
      wherein one should set $a_0\geq 0$ and $a_1<0$ in order to exclude
      tachyons~\cite{goms}.

      \indent

      Now, a point is that: \emph{the HDN provides an easy procedure to classify different
      Lagrangian terms, especially higher order Lagrangians.}

      \indent

      For example, in order to amend the Lagrangian of sixth order
      gravity,
      Berkin {\it et al}~\cite{berm}\ discussed that the Lagrangian term of
      $R\,\Square\, R$ is a \emph{third} order Lagrangian based on the
      dimensionality scale, i.e. two derivatives are dimensionally
      equivalent to one Riemann--Christoffel tensor or any one of its
      contractions. However, it can be better justified on account of the
      above regard, since it has the HDN {\it three}~\cite{farb}. Hence, in order to
      generalize Lagrangian~(\ref{sixo}), the following
      Lagrangian has also been considered in the literature~\cite{berm,goms}
      \begin{equation}\label{lsixo}
      L={1\over \kappa^2}\Bigl(R+a_0\, R^2+b\, R^3
        +a_1\, R\, \Square\, R\Bigr)\ ,
      \end{equation}
      where $b$ is a constant of the order of (Planck's) length to the
      power 4. Later on, the Lagrangian
      \begin{equation}\label{gsixo}
      L={1\over \kappa^2}\Bigl[R+a_0\, R^2
        +a_N\, \bigl(-R\bigr)^N\,\Square\, R\Bigr]\ ,
      \end{equation}
      where $N$ is a positive integer and $a_N$ is a constant of the order
      of (Planck's) length to the power $2N+2$, has been regarded
      as generalized sixth order gravity~\cite{aetal}\rlap.\footnote{The
       sign convention is such that the de Sitter space--time has a negative
       curvature value~\cite{aetal}. }\
      Obviously, the HDN of $(-R)^N\,\Square\, R$ is $N+2$ \cite{farb}, and hence,
      Lagrangian~(\ref{gsixo}) does~not contain all the terms constructed from the
      curvature scalar with equal HDNs, e.g. see below.

      As another example, in addition to the eight linearly
      independent terms which appear in the third order of the Lovelock
      Lagrangian and are up to the $2^{nd}$ order jet-prolongation of the
      metric with the HDN {\it three}, namely~\cite{mulhc,fard,whelt}
      {\setlength\arraycolsep{2pt}
      \begin{eqnarray}
      &R^3
      \qquad RR_{\mu\nu}R^{\mu\nu}
      \quad  &RR_{\rho\tau\mu\nu}R^{\rho\tau\mu\nu}
      \qquad R^{\mu\nu}R_{\mu\gamma}R_{\nu}{}^{\gamma}
      \qquad R_{\rho\tau}R_{\mu\nu}R^{\rho\mu\tau\nu}\cr
      &R_{\lambda\rho}R^{\lambda\tau\mu\nu}R^{\rho}{}_{\tau\mu\nu}
      \qquad &R^{\sigma\tau}{}_{\mu\nu}R^{\mu\nu}{}_{\lambda\rho}
                                  R^{\lambda\rho}{}_{\sigma\tau}
      \qquad  R^{\sigma\tau}{}_{\mu\nu}R^{\mu\lambda}{}_{\sigma\rho}
                           R^{\nu\rho}{}_{\tau\lambda}\ ,
      \end{eqnarray} }there are generally another
      nine\footnote{However, $\Square\,^2 R$ is a complete divergence and has no
       effect in the variation of the action. Besides,
       not all of their corresponding Euler--Lagrange expressions
       are independent~\cite{farf}, e.g. $R_{;\,\rho}R^{;\,\rho}$, upon
       integration covariantly by parts, can be transferred to a boundary term
       plus the $R\,\Square\, R$ term. }\
      linearly independent scalar terms constructed from the
      Riemann--Christoffel tensor and its contractions which can have the
      HDN of {\it three} as well, though they are up to $3^{rd}$ or even
      higher order jet-prolongation of the metric, namely
      {\setlength\arraycolsep{2pt}
      \begin{eqnarray}
      &R\,\Square\, R
      \,\,\quad &R_{\mu\nu}\,\Square\, R^{\mu\nu}
      \qquad R_{\mu\nu\rho\tau}\,\Square\, R^{\mu\nu\rho\tau}
      \qquad R_{\mu\nu}R^{;\,\mu\nu}
      \qquad R_{\mu\nu ;\,\rho}R^{\mu\rho ;\,\nu}\cr
      &\Square\,^2 R
      \,\,\quad &R_{;\,\rho}R^{;\,\rho}
      \,\quad\qquad R_{\mu\nu ;\,\rho}R^{\mu\nu ;\,\rho}
      \,\quad\qquad R_{\mu\nu\rho\tau ;\,\alpha}R^{\mu\nu\rho\tau ;\,\alpha}\ .
      \end{eqnarray} }The higher order terms become increasingly complex, e.g. the
      full expression for the fourth order of the Lovelock
      Lagrangian which are up to the $2^{nd}$ order jet-prolongation of the
      metric with the HDN {\it four} has 25 terms~\cite{brig,whelt,defa}.

\item The second term of the Lovelock Lagrangian, the Gauss--Bonnet term,
      does indeed satisfy the condition~(\ref{cnlii}). In another words, the
      coefficients used in the Lanczos Lagrangian are specific to a equivalence
      class of any generic second order Lagrangians accepting our
      analogous demand.

\item If the constraint~(\ref{cnlii}) is applied to the Lagrangian made from
      the square of the Weyl conformal tensor, namely
      \begin{equation}\label{wsd}
      C_{\alpha\beta\mu\nu}\,C^{\alpha\beta\mu\nu}=
      {2\over(D-1)(D-2)}R^2-{4\over(D-2)}R_{\mu\nu}R^{\mu\nu}+
      R_{\alpha\beta\mu\nu}\,R^{\alpha\beta\mu\nu}\ ,
      \end{equation}
      gives
      \begin{equation}
      3\, a_1+a_2+a_3={6\over(D-1)(D-2)}-{4\over(D-2)}+1
      ={(D-3)(D-4)\over(D-1)(D-2)} ,
      \end{equation}
      which can only be zero when the dimension of space--time is either
      \emph{three} or \emph{four}. Though, in three dimension, the Weyl
      tensor is itself identically zero. And, as the action constructed
      by the square of the Weyl tensor, i.e. $I\equiv \int
      C_{\alpha\beta\mu\nu}C^{\alpha\beta\mu\nu}\sqrt{-g}\> d^D\!x$,
      conformally transforms as $I\to \Omega^{D-4}\,I$,
      it is thus conformal invariant only in {\it four}
      dimensions\rlap.\footnote{Actually, in four dimensions
       the only local geometrical conformal invariant that can be
       constructed from the metric tensor and its first and second
       derivatives is
       $C_{\alpha\beta\mu\nu}C^{\alpha\beta\mu\nu}\sqrt{-g}$.}\

      Hence, following the classification of Ref.~\cite{desch},
      the constraint~(\ref{cnlii}) indicates that in \emph{four}
      dimensions, the anomaly can have two contributions, a type
      $A$ anomaly, i.e. the Euler density invariant, and a type
      $B$ anomaly built from conformal invariants, i.e. the Weyl
      squared in four dimensions.

\item The condition~(\ref{cnlii}) does~not permit the
      inclusion of only one of the second order Lagrangians in
      isolation, as mentioned before, difficulties also exist in quantum gravity
      when only one of these second order terms is considered~\cite{stelb,stela}.

\item As the most general quadratic Lagrangian in (and up to) four dimensions
      is~(\ref{eplii}), the condition~(\ref{cnlii}) reads
      \begin{equation}\label{cnliif}
      3\,\alpha+\beta=0\ ,
      \end{equation}
      where $\alpha$ and $\beta$ are arbitrary non-zero constants.

      In this case, the particle spectrum in the quantum theory does~not contain
      the scalar particle, though this does~not help the unitarity
      problems~\cite{stelb}.

\item In two dimensions, as $R_{\alpha\beta\mu\nu}\,
      R^{\alpha\beta\mu\nu}=2 R_{\mu\nu}R^{\mu\nu}=R^2$, there is only
      one independent second order Lagrangian that can be considered as
      a higher derivative term~\cite{hwkvyon}. However,
      condition~(\ref{cnlii}) does~not allow it, just as one cannot have
      Einstein's Lagrangian term in two dimensional space--time either
      and other options have been sought~\cite{bro}.

\item To enquire about the cosmological term, let us investigate the
      analogy for this as well. Consistent with the aspect of
      equations~(\ref{love}) and~(\ref{lovet}) for the $n=0$ case (zero
      HDN), the Lagrangian $c_0\, L^{(0)}\equiv 2\Lambda/\kappa^2$, a
      constant, produces the cosmological term,
      $G^{(0)}_{\mu\nu}=-\Lambda\, g_{\mu\nu}$, in the field equations.
      Hence, the exception value of the HDN in our definition of
      generalized trace\rlap,\footnote{That is, e.g. for any
       tensor $T_{\mu\nu}$ with HDN of $h$, we set ${\sl Trace}\,
       {}^{[h]}T_{\mu\nu}={1\over h+1}\, {\sl trace}\,
       T_{\mu\nu}$ when $h\not=-1$, see Ref.~\cite{farb} for details.}\
      equation~(\ref{Tracedu}), maybe related to the
      cosmological term difficulty.
      Nevertheless, with our choice of definition for the generalized
      trace, we have ${\sl Trace}\, {}^{[+1]}g^{\mu\nu}={\sl trace}\,
      g^{\mu\nu}=D$ and ${\sl Trace}\, {}^{[-1]}g_{\mu\nu}={\sl
      trace}\, g_{\mu\nu}=D$, as though the dimension of space--time
      has~not been altered.

      Now, if $D\neq 2$, one may, for example, set
      \begin{equation}\label{cosmo}
      G^{(0)}_{\mu\nu}=\bigl[{2\Lambda\over D-2}\,
      g_{\mu\nu}\bigr]-{1\over 2}g_{\mu\nu}\bigl[{2D\Lambda\over
      D-2}\bigr]\equiv R^{(0)}_{\mu\nu}-{1\over 2}g_{\mu\nu}R^{(0)}\ ,
      \end{equation}
      which holds in the trace relation, i.e. ${\sl Trace}\,
      R^{(0)}_{\mu\nu}=R^{(0)}$. Therefore, the inclusion of the
      cosmological term is allowed by our analogous demand.

\item As final remark, we claim that the analogous demand, that we
      have introduced, provides a more compatible understanding
      with the Mach idea. Let us first elaborate a few words on
      the background of this issue.

      Stachel \cite{sta} has pointed out that what went wrong in
      the last conclusion of the so-called Einstein's ``hole'' argument
      was that the point events of the space--time manifold had been
      incorrectly thought of as \emph{individuated} independently of the
      field itself. That is, it is impossible to drag the metric field
      away from a physical point in empty space--time and leave that
      physical point behind. Actually, Einstein himself
      realized~\cite{einleinll} this as he wrote to Besso that
      nothing is physically \emph{real} but the \emph{totality} of
      space--time point coincidences. Later on, in an
      addendum~\cite{eindd} he placed great stress on the
      inseparability of the metric and the manifold.

      Besides, according to the Aristotelian view~\cite{adl}, space is
      a \emph{plenum}, i.e. it is inseparably associated with the material
      substance, and not a \emph{void}. Hence, the properties of space
      are~not independent of the material bodies that move \emph{in}
      it. Just as it is well known that, even though in
      Newtonian physics, space is a pre-existing stage on which material
      particles are the characters acting out the drama of physical
      events, but, in relativistic gravitational physics, space cannot
      be considered apart from the matter that is in it. And, as the
      mathematician E. Whittaker points out, the characters create the
      stage as they walk about on it~\cite{adl}. Actually, as it has been
      pointed out~\cite{mashliuwess94}, a basic problem of Newtonian
      mechanics is that the extrinsic state of a point particle, i. e.
      its appearance in space and time (usually characterized by its
      position and velocity), is a \emph{priori} independent of its
      intrinsic state (usually characterized by its mass). Hence, one may
      conclude that the properties of space--time in gravitational theories
      must be inseparable from the matter that is in it.

      On the other hand, though the main critique of Mach and Einstein to
      absolute space, that it acts on everything but is~not affected by
      anything, is~not applied to GR, but,
      actually, according to the Mach idea, the question would be
      whether matter just modifies an already existing space--time
      structure, or whether it is the only source for its
      structure~\cite{gra}. The former, a weaker version of Mach's principle,
      is in agreement with Einstein's gravitational theory. But, the
      latter, a strong version of Mach's principle in the sense that for
      a universe devoid of matter there should be no meaning for the
      existence of space--time~\cite{gra}, is~not consistent with
      it, for it has specified structures in the absence of
      matter\rlap,\footnote{See, for example Ref. \cite{rysh}.}\
      e.g. the Minkowski, Schwarzschild and Taub--NUT metrics~\cite{schntu}.
      So that, the space--time of GR
      still itself has some essence independent of matter.

      The link to this issue, in our opinion, could be the relation
      between $T$ and $R$ as in ${\sl Trace}\,
      G_{\alpha\beta}=\bigl(1-{D\over 2}\bigr)R
       ={1 \over 2}\kappa^2\, T$, and the basic concept of matter.
      For example, a similar behavior, as $R$ vanishes whenever
      $T_{\alpha\beta}$ vanishes, should also be sought for the higher
      order terms. This
      is somehow a procedure that it may indicate a more compatibility
      between GR and the strong version of the Mach
      idea~\cite{abbasi}. In another words, we adopt the view for establishing
      the generality of the known proposals, namely, that
      geometrical curvature induces matter~\cite{sal}, geometrical
      description of physical forces, and geometrical origin for
      the matter content of the universe. As the idea, more or less
      in this connection, that associates extra dimensions with the intrinsic
      characteristics of matter, e.g.~\cite{mashliuwess94}, or
      any analytical $D$ dimensional Riemannian manifold can be locally
      embedded in an $(D+1)$ dimensional flat Riemannian
      manifold~\cite{rtz}, see Refs.~\cite{wesson99and2005} for a more
      extensive discussion.

      Hence, using the analogous demand, we showed~\cite{farc} that in
      the cases where higher order gravities dominate, space--time
      ``behaves'' as if its energy-momentum has been ``exchanged'' with
      matter's energy-momentum in the sense that in a universe devoid of
      ``matter'' there should be also no meaning for the existence of
      space--time. Note that, the applicability of higher order
      gravitational theories are restricted by the energy scale. In
      other words, as the coefficients of higher order gravities are
      very small, one cannot detect such implications in ``real''
      world. However, these effects are important in highly curved
      areas, such as the very early universe, or in quantum physics.
\end{itemize}

\section{Dimensional Dependent Version of Duff's Relation}
\indent

The appearance of $G\!^{(2)}_{({\rm generic})\alpha\beta}$ and the
order through which one can define $R\!^{(2)}_{({\rm
generic})\alpha\beta}$ and $R\!^{(2)}_{\rm generic}$ is somehow a
critical point. In this section, we order/factorize the
equation~(\ref{nlii}) in an alternative and more basic approach,
which is commonly employed in the process of varying the
action~\cite{farb}, namely
 {\setlength\arraycolsep{2pt}
\begin{eqnarray}\label{nliia}
&G&\!^{(2)}_{({\rm generic})\alpha\beta}\! \equiv \!{\kappa^2\over
   \sqrt{-g}}\,{\delta\bigl(L^{(2)}_{\rm generic}\sqrt{-g}\bigr)\over
   \delta g^{\alpha\beta}}=\kappa^2\,{\delta\bigl(L^{(2)}_{\rm
   generic}\bigr)\over \delta g^{\alpha\beta}}-{1\over 2}\kappa^2\,
   g_{\alpha\beta}\, L^{(2)}_{\rm generic}\cr
&=&2\biggl[ a_1 R\,R_{\alpha\beta}-2a_3
   R_{\alpha\mu}\,R_{\beta}{}^{\mu}+a_3
   R_{\alpha\rho\mu\nu}\,R_{\beta}{}^{\rho\mu\nu}+\bigl(a_2+2a_3\bigr)
   R_{\alpha\mu\beta\nu}\,R^{\mu\nu}\cr
&&{}\quad -\bigl(a_1+{1\over
   2}a_2+a_3\bigr)\,R_{;\,\alpha\beta}+\bigl({1\over
   2}a_2+2a_3\bigr)\,\Square\, R_{\alpha\beta}+{1\over
   4}\bigl(4a_1+a_2\bigr)\times\cr
&&{}\qquad g_{\alpha\beta}\,\Square\, R\biggr]
   -{1\over 2}\kappa^2\, g_{\alpha\beta}\, L^{(2)}_{\rm generic}\cr
&&{} \equiv R^{(2^\prime)}_{({\rm generic}) \alpha\beta}
   -{1\over 2}g_{\alpha\beta}\, R^{(2^\prime)}_{\rm generic}\ .
\end{eqnarray} }
As it is evident, in this case, $R^{(2^\prime)}_{({\rm generic})
\alpha\beta}\equiv \kappa^2\,{\delta\bigl(L^{(2)}_{\rm
   generic}\bigr)\over \delta g^{\alpha\beta}}$ and
$R^{(2^\prime)}_{\rm generic}\equiv \kappa^2\, L^{(2)}_{\rm
generic}$ are completely similar to the case of GR. This is also a
more familiar manner which has been used by Lovelock~\cite{lovedc}
for the non-generic case, though he then proceeded from this to
derive equation~(\ref{lovet}).

The ``trace'' relation, ${\sl Trace}\, R^{(2^\prime)}_{({\rm
generic})\alpha\beta}=R^{(2^\prime)}_{\rm generic}$, holds if and
only if, for non~zero coefficients,
\begin{equation}\label{cnliia}
 \Boxed{ \bigl(D-1\bigr)a_1+{D\over 4}a_2+a_3=0}
       \qquad {\rm for}\ D\geq 2\ .
\end{equation}
As it is evident, the explicit appearance of $D$ gives rise to a
dimensional dependent version of Duff's trace anomaly relation.
Though, one may argue that usually the trace anomaly in $2D$
dimensions is related to $R^D$--like invariants, i.e. the $D$
order Lagrangian terms. Actually, investigation
shows~\cite{fard,farf}\ that similar constraint relations for the
third order Lagrangian terms hold which Weyl invariants in six
dimensions~\cite{bopb,desch}\ do satisfy them.

Again, the constraint~(\ref{cnliia}) gives two degrees of freedom,
in $D > 4$, to choose a desired combination. Also, only in
\emph{four} dimensions, it is exactly the same as the
condition~(\ref{cnlii}), where, due to the extra condition of the
Gauss--Bonnet density, leads again in this dimension to the
condition~(\ref{cnliif}). Hence, the constraint has been modified
for higher dimensions. The reason is obvious, for that one can
generally set {\setlength\arraycolsep{2pt}
\begin{eqnarray}
{}&{}&R'_{\alpha\beta}\equiv
      R^{(2)}_{(\rm generic)\alpha\beta}-g_{\alpha\beta}\,P\cr
    &&R'\equiv R^{(2)}_{\rm generic}-2P\ ,
\end{eqnarray} }
where $P$ can be an arbitrary homogeneous function of any order
jet-prolongation of the metric with the HDN {\it two}. But,
using~(\ref{Tracexampl}), one gets
\begin{equation}
{\rm Trace}\,R'_{\alpha\beta}=R^{(2)}_{\rm generic}-{D\over 2}\,
P\ .
\end{equation}
And, in order to be equal to $R'$, $D$ must be four. In the
special case of $R^{(2^\prime)}_{(\rm generic)\alpha\beta}$ and
$R^{(2^\prime)}_{\rm generic}$, $P$ is obviously proportional to
$\,\Square\, R$. Incidentally, since the term $\,\Square\, R$ is a
complete divergence, it gives no contribution to the variation of
the relevant action. Therefore, adding such a term to the
Lagrangian still gives the same result, but it can cover the
remaining part of the trace anomaly~(\ref{fanom}).

The non-generic combination of the Lanczos Lagrangian satisfies
the above constraint in \emph{any} dimension. Actually, one can
write equation~(\ref{cnliia}) as
\begin{equation}
D\bigl(a_1+{1\over 4}a_2\bigr)+\bigl(a_3-a_1\bigr)=0\ ,
\end{equation}
which shows that the \emph{only} combination that is independent
of dimension is the combination of the Lanczos Lagrangian.

The square of the Weyl conformal tensor, equation~(\ref{wsd}),
identically satisfies the condition~(\ref{cnliia}) independent of
dimension. Hence, the modified derivation of the ``trace''
relation yields equation~(\ref{cnliia}) which now does give the
Euler density invariant and the Weyl squared in \emph{arbitrary}
dimensions.

To reiterate, if one considers only one of the second order
Lagrangians~(\ref{lii}) in isolation, constraint~(\ref{cnliia})
will set the third coefficient equal to zero as well. In three
dimensions, again because of the extra Gauss--Bonnet relation, the
constraint, for the Lagrangian~(\ref{eplii}), now reads
\begin{equation}\label{cnliiat}
{8\over 3}\alpha+\beta=0\ .
\end{equation}
In this case, the particle spectrum in the quantum theory should
be rechecked~\cite{farh}. In two dimensions, the constraint gives
\begin{equation}\label{cnliibt}
a_1+{1\over 2}a_2+a_3=0\ ,
\end{equation}
where, because of the relation mentioned before, the Lagrangian
under consideration actually is $\bigl(a_1+{1\over
2}a_2+a_3\bigr)R^2$. Hence, the above constraint, in two
dimensions, again confirms the null result of the previous
section. But, the arrangement~(\ref{cosmo}), for the cosmological
term, cannot be performed.

Finally, the Euler--Lagrange expression $G^{(2)}_{({\rm
generic})\alpha\beta}$ could partially be disarranged and be
written as
{\setlength\arraycolsep{2pt}
\begin{eqnarray}\label{nliib}
 &&G^{(2)}_{({\rm generic})\alpha\beta}=\cr
             &&2\biggl\{a_1 R\, R_{\alpha\beta}
               -2a_3 R_{\alpha\mu}\, R_{\beta}
               {}^{\mu}+a_3 R_{\alpha\rho\mu\nu}\, R_{\beta}{}^{\rho\mu\nu}
               +\bigl(a_2+2a_3\bigr)R_{\alpha\mu\beta\nu}\,
               R^{\mu\nu}-\bigl(a_1\cr
             &&\quad +{1\over 2}a_2+a_3\bigr)\,
               R_{;\,\alpha\beta}
               +\bigl({1\over 2}a_2+2a_3\bigr)\,
               \Square\, R_{\alpha\beta}
               +a\Bigl[{1\over 4}\bigl(4a_1+a_2\bigr)+a_3\Bigr]
               g_{\alpha\beta}\,\Square\, R\biggr\}\cr
             &&-{1\over 2}\, g_{\alpha\beta}\biggl\{\kappa^2\,
               L^{(2)}_{\rm generic}+\Bigl[\bigl(4a_1+a_2\bigr)
               \bigl(a-1\bigr)+4\, a\, a_3\Bigr]\,\Square\, R
               \biggr\}\ ,
\end{eqnarray} }
where $a$ is a number. In this case, the ``trace'' requirement
holds if and only if
\begin{equation}\label{cnliib}
a_1\Bigl[a\bigl(D-4\bigr)+3\Bigr]+a_2\Bigl[{a\over
4}\bigl(D-4\bigr)+1\Bigr]
      +a_3\Bigl[a\bigl(D-4\bigr)+1\Bigr]=0\ .
\end{equation}
The most interesting point of the above condition is that, in four
dimensions it reduces to condition~(\ref{cnlii}) irrespective of
the ``disarrangement'' factor $a$.

\section*{Acknowledgement}
\indent

This work has been supported by the research council of the Shahid
Beheshti University, Tehran, Iran.



\begin{thebibliography}{9}
\bibitem{ishishb}Isham, C J ``Quantum gravity'' in Proc. 11$^{th}$
                 {\it General Relativity and Gravitation}, Stockholm,
                 1986, Ed. M A H MacCallum (Cambridge University
                 Press, 1987), pp 99-129; ``Structural issues in quantum
                 gravity'', {\it gr-qc/9510063}.
\bibitem{negogoefran}Newman, E T \& Goenner, H ``Classical and quantum
                     alternatives to gravitational theories'' in Proc.
                     10$^{th}$ {\it General Relativity and Gravitation},
                     Padua, Italy, 1983, Eds. B Bertotti, F de Felice \&
                     A Pascolini (D Reidel Publishing Company, Holland,
                     1984), pp 199-211;\newline
                     Goenner, H ``Alternative theories of gravity'' in
                     Proc. 11$^{th}$ {\it General Relativity and Gravitation},
                     Stockholm, 1986, Ed. M A H MacCallum (Cambridge University
                     Press, 1987), pp 262-273;\newline
                     Francaviglia, M ``Alternative gravity theories'' in Proc.
                     12$^{th}$ {\it General Relativity and Gravitation},
                     Boulder, 1989, Eds. N Ashby, D F Bartlett \& W Wyss
                     (Cambridge University Press, 1990), pp 99-104.
\bibitem{hil}Hilbert, D ``Die Grundlagen der Physik'', I \& II,
             {\it Nachr. Gesel. Wiss. G\"ottingen}, (1915), 395-407 and
             (1917), 53-76, respectively. These were consolidated with the
             same title into: {\it Math. Annalen}\ {\bf 92}\ (1924), 1-32.
\bibitem{weyl}Weyl, H ``Gravitation und Elektrizit\"at'',
              {\it Preuss. Akad. Wiss. Berlin, Sitz.}\ (1918), 465-480;
              ``Eine neue Erweiterung der Relativit\"atstheorie'',
              {\it Ann. der Phys.}\ {\bf 59}\ (1919), 101-133;
              {\it Raum-Zeit-Materie}, (Springer-Verlag, Berlin, 1$^{st}$
              ed.~1918, 4$^{th}$ ed.~1921). Its English version
              (of the 4$^{th}$ ed.) is: {\it Space--Time--Matter},
              translated by: H L Brose (Dover Publications, New York,
              1$^{st}$ ed.~1922, reprinted 1950);
              ``\"Uber die physikalischen Grundlagen der erweiterten
              Relativit\"atstheorie'', {\it Phys. Zeitschr.}\ {\bf 22}\
              (1921), 473-480.
\bibitem{edd}Eddington, A, {\it The Mathematical Theory of Relativity},
             (Cambridge University Press, 1$^{st}$ ed.~1923, 2$^{nd}$
             ed.~1924; Chelsee Publishing Co., New York, 1975).
\bibitem{stelb}Stelle, K S ``Classical gravity with higher derivatives'',
               {\it \GRG}\ {\bf 9}\ (1978), 353-371.
\bibitem{malu}Maluf, W ``Conformal invariance and torsion in general
              relativity'', {\it \GRG}\ {\bf 19}\ (1987), 57-71.
\bibitem{bac}Bach, R ``Zur Weylschen Relativit\"atstheorie und der
             Weylschen Erweiterung des Kr\"ummungstensorbegriffs'',
             {\it Math. Zeitschr.}\ {\bf 9}\ (1921), 110-135.
\bibitem{lanalanaa}Lanczos, C ``Elektromagnetismus als nat\"urliche
                   Eigenschaft der Riemanns chen Geometrie'',
                   {\it Zeits. Phys.}\ {\bf 73}\ (1931), 147-168;
                   ``Electricity and general relativity'',
                   {\it \RMP}\ {\bf 29}\ (1957), 337-350.
\bibitem{buchdcbick}Buchdahl, H A ``On the gravitational field equations
                    arising from the square of a Gaussian curvature'',
                    {\it Nuovo Cim.}\ {\bf 23}\ (1962), 141-157;\newline
                    Bicknell, G V ``Non-viability of gravitational theory
                    based on a quadratic Lagrangian'', {\it J. Phys. A:
                    Math. Nucl. Gen.}\ {\bf 7} (1974), 1061-1069.
\bibitem{cliftonba}Clifton, T \& Barrow, J D ``The power of
                   general relativity'', {\it gr-qc/0509059}.
\bibitem{barrowcl}Barrow, J D \& Clifton, T ``Exact cosmological
                  solutions of scale--invariant gravity theories'',
                  {\it gr-qc/0509085}.
\bibitem{buch70}Buchdahl, H A ``Non--linear Lagrangians and
                cosmological theory'', {\it Mon. Not. R. Astr. Soc.}\
                {\bf 150}\ (1970), 1-8.
\bibitem{wheb}Wheeler, J A, {\it Einstein's Vision}, (Springer-Verlag,
              Berlin, 1968).
\bibitem{ishb}Isham, C J ``Structural issues in quantum gravity'',
              {\it gr-qc/9510063}, (lectures delivered at the GR14
              Conference, Florence, August 1995).
\bibitem{cfgpp}Cecotti, S, Ferrara, S, Girardello, L, Porrati, M \&
               Pasquinucci, A ``Matter coupling in higher derivative
               supergravity'', {\it \PR}\ {\bf D33}\ (1986), R2504-R2507.
\bibitem{bidabos}Birrell, N D \& Davies, P C W, {\it Quantum Fields in Curved
                 Space}, (Cambridge University Press, 1982);\newline
                 Buchbinder, I L, Odintsov, S D \& Shapiro, I L, {\it
                 Effective Action in Quantum Gravity}, (Institute of Physics
                 Publishing Bristol and Philadelphia, 1992).
\bibitem{utdepese}Utiyama, R \& DeWitt, B S ``Renormalization of a
                  classical gravitational interacting with quantized
                  matter fields'', {\it \JMP}\ {\bf 3} (1962),
                  608-618;\newline
                  Pechlaner, E \& Sexl, R ``On quadratic Lagrangians in
                  general relativity'', {\it Commun. Math. Phys.}\
                  {\bf 2}\ (1966), 165-175.
\bibitem{cheachebkonospiv}Chern, S-S ``A simple intrinsic proof of the
                          Gauss--Bonnet formula for closed Riemannian
                          manifolds'', {\it \AM}\ {\bf 45}\ (1944), 747-752;
                          ``On the curvatura integra in a Riemannian
                          manifold'', {\it ibid.}\ {\bf 46}\ (1945),
                          674-684;\newline
                          Kobayashi, S \& Nomizu, K, {\it Foundations of
                          Differential Geometry}, Vol.~II, (Wiley
                          Interscience, New York, 1969);\newline
                          Spivak, M, {\it A Comprehensive Introduction to
                          Differential Geometry}, Vol.~5, (Publish or
                          Perlish Inc., Delaware, 2$^{nd}$ ed.~1979).
\bibitem{stela}Stelle, K S ``Renormalization of higher derivative quantum
               gravity'', {\it \PR}\ {\bf D16}\ (1977), 953-969.
\bibitem{frtsa}Fradkin, E S \& Tseytalin, A A ``Renormalizable
               asymptotically free quantum theory of gravity'',
               {\it \NP}\ {\bf B201}\ (1982), 469-491.
\bibitem{chswmtgswltkp}Candelas, P, Horowitz, G T, Strominger, A \& Witten, E
                       ``Vacuum configurations for superstrings'',
                       {\it \NP}\ {\bf B258}\ (1985), 46-74;\newline
                       Metsaev, R R \& Tseytlin, A A ``Curvature cubed terms in
                       string theory effective actions'', {\it \PL}\ {\bf B185}\
                       (1987), 52-58;\newline
                       Green, M B, Schwarz, J H \& Witten, E, {\it Superstring
                       Theory}, Vols.~1 \&~2, (Cambridge University Press,
                       1987);\newline
                       Lust, D \& Theusen, S, {\it Lectures on String Theory},
                       (Springer, Berlin, 1989);\newline
                       Ketov, S V ``The string generated correction to Einstein
                       gravity from the sigma model approach'', {\it \GRG}\
                       {\bf 22}\ (1990), 193-202;\newline
                       Polchinski, J, {\it String Theory}, (Cambridge
                       University Press, 1998).
\bibitem{dese}Deser, S ``Gravity from strings'', {\it  Phys. Scripta}\
              {\bf T15}\ (1987), 138-142.
\bibitem{lovedc}Lovelock, D ``The Einstein tensor and its
                generalizations'', {\it \JMP}\ {\bf 12}\ (1971), 498-501;
                ``The four dimensionality of space and the Einstein
                tensor'', {\it ibid.}\ {\bf 13}\ (1972), 874-876.
\bibitem{brig}Briggs, C C ``Some possible features of general expressions for
              Lovelock tensors and for the coefficients of Lovelock
              Lagrangians up to the $15^{th}$ order in curvature (and beyond)'',
              {\it gr-qc/9808050}.
\bibitem{wald}Wald, R M, {\it General Relativity}, (The University of
              Chicago Press, 1984).
\bibitem{zwizum}Zwiebach, B ``Curvature squared terms and string
                theories'', {\it \PL}\ {\bf B156}\ (1985),
                315-317;\newline
                Zumino, B ``Gravity theories in more than four
                dimensions'', {\it \PRp}\ {\bf 137}\ (1986), 109-114.
\bibitem{dnpb}Duff, M J, Nilsson, B E W \& Pope, C N ``Gauss--Bonnet
              from Kaluza--Klein'', {\it \PL}\ {\bf B173}\ (1986), 69-72.
\bibitem{chammuh}Chamseddine, A H ``Topological gauge theory of gravity
                 in five and all odd dimensions'', {\it \PL}\ {\bf B233}\
                 (1989), 291-294;\newline
                 M\"uller-Hoissen, F ``From Chern--Simons to
                 Gauss--Bonnet'', {\it \NP}\ {\bf B346}\ (1990), 235-252.
\bibitem{allfrr}Allemandi, G, Francaviglia, M \& Raiteri, M
                ``Charges and energy in Chern--Simons theories and
                Lovelock gravity'', {\it \CQG}\ {\bf 20}\ (2003),
                5103-5120.
\bibitem{nood}Nojiri, S \& Odintsov, S D ``Where new gravitational physics
              comes from M--theory?'', {\it Phys. Lett.}\ {\bf B576}\
              (2003), 5-11.
\bibitem{simo}Simon, J Z ``Higher derivative Lagrangians, non-locality,
              problems, and solutions'', {\it \PR}\ {\bf D41}\ (1990),
              3720-3733.
\bibitem{mdmmmpcf}Madore, J ``Cosmological applications of the Lanczos
                  Lagrangian'', {\it \CQG}\ {\bf 3}\ (1986),
                  361-371;\newline
                  Deruelle, N \& Madore, J ``Kaluza--Klein cosmology with
                  the Lovelock Lagrangian'' in {\it Origin and Early
                  History of the Universe}, Proc.~26$^{th}$
                  {\it Li\`ege Int. Astrophysical Colloquium}, Belgium, July
                  1986, (Cointe-Ougree, Belgium, 1987), pp
                  277-283;\newline
                  Mena Marug\'an, G A ``Classical and quantum
                  Lovelock cosmology'', {\it \PR}\ {\bf D42}\ (1990),
                  2607-2620; ``Lovelock gravity and classical
                  wormholes'', {\it \CQG}\ {\bf 8}\ (1991),
                  935-946;\newline
                  Poisson, E ``Quadratic gravity and the black hole
                  singularity'', {\it \PR}\ {\bf D43}\ (1991), 3923-3928;\newline
                  Cotsakis, S, {\it Cosmological Models in Higher Order Gravity},
                  (Ph.D. Thesis, Sussex University, 1990);\newline
                  Fari\~na-Busto, L, {\it Non--Linear Gravitational Lagrangians
                  in Cosmology}, (Ph.D. Thesis, Queen Mary \& Westfield College,
                  University of London, 1990).
\bibitem{cdct}Carloni, S, Dunsby, P K S, Capozziello, S \& Troisi,
              A ``Cosmological dynamics of $R^n$ gravity'', {\it \CQG}\ {\bf 22}\
              (2005), 4839-4868.
\bibitem{berm}Berkin, A L \& Maeda, K ``Effects of $R^3$ and
              $R\,\Square\, R$ terms on $R^2$ inflation'', {\it \PL}\
              {\bf B245}\ (1990), 348-354.
\bibitem{schmcbermb}Schmidt, H-J ``Variational derivatives of arbitrarily high
                    order and multi-inflation cosmological models'',
                    {\it \CQG}\ {\bf 7}\ (1990), 1023-1031;\newline
                    Berkin, A L \& Maeda, K ``Inflation in
                    generalized Einstein theories'', {\it \PR}\ {\bf D44}\
                    (1991), 1691-1704.
\bibitem{goms}Gottl\"ober, S, M\"uller, V \& Schmidt, H-J ``Generalized
              inflation from $R^3$ and $R\,\Square\, R$ terms'',
              {\it  Astron. Nachr.}\ {\bf 312}\ (1991), 291-297.
\bibitem{dehamengwangabf}Dehghani, M H ``Magnetic branes in Gauss--Bonnet gravity'',
                         {\it \PR}\ {\bf D69}\ (2004), 064024;\newline
                         Meng, X-H, \& Wang, P ``Inflationary
                         attractor in Gauss--Bonnet brane cosmology'',
                         {\it \CQG}\ {\bf 21}\ (2004), 2527-2536;\newline
                         Allemandi, G, Borowiec, A \&
                         Francaviglia, M ``Accelerated
                         cosmological models in Ricci squared
                         gravity'', {\it \PR}\ {\bf D70}\ (2004),
                         103503.
\bibitem{dehb}Dehghani, M H ``Accelerated expansion of the universe in
              Gauss--Bonnet gravity'', {\it Phys. Rev.}\ {\bf D70}\
              (2004), 064009.
\bibitem{rptetal}Riess, A G, {\it et al}\ ``Observational evidence from supernovae for
                 an accelerating universe and a cosmological constant'', {\it Astro. J.}\
                 {\bf 116}\ (1998), 1009-1038;\newline
                 Perlmutter, S, {\it et al}\ ``Measurements of omega and lambda from
                 42 high-redshift supernovae'', {\it Astrophys. J.}\ {\bf 517}\ (1999),
                 565-586;\newline
                 Tonry, J L, {\it et al}\ ``Cosmological results from high-z
                 supernovae'', {\it Astrophys. J.}\ {\bf 594}\ (2003), 1-24.
\bibitem{lnhsetal}Lee, A T, {\it et al}\ ``A high spatial resolution analysis of the
                  MAXIMA-1 cosmic microwave background anisotropy data'',
                  {\it Astrophys. J.}\ {\bf 561}\ (2001), L1-L6;\newline
                  Netterfield, C B, {\it et al}\ ``A measurement by BOOMERANG of
                  multiple peaks in the angular power spectrum of the cosmic
                  microwave background'', {\it Astrophys. J.}\ {\bf 571}\ (2002),
                  604-614;\newline
                  Halverson, N W, {\it et al}\ ``DASI first results: a measurement of
                  the cosmic microwave background angular power spectrum'', {\it
                  Astrophys. J.}\ {\bf 568}\ (2002), 38-45;\newline
                  Spergel, D N, {\it et al}\ ``First year Wilkinson Microwave
                  Anisotropy Probe (WMAP) observations: determination of cosmological
                  parameters'', {\it Astrophys. J. Suppl.}\ {\bf 148}\ (2003), 175-225.
\bibitem{cdttccct}Carroll, S M, Duvvuri, V, Trodden, M \& Turner, M S
                  ``Is cosmic speed-up due to new gravitational
                  physics?'', {\it Phys. Rev.}\ {\bf D70}\ (2004), 043528;\newline
                  Capozziello, S, Cardone, V F, Carloni, S \& Troisi, A ``Curvature
                  quintessence matched with observational data'', {\it Int. J. Mod.
                  Phys.}\ {\bf D12} (2003), 1969-1982; ``Higher order curvature
                  theories of gravity matched with observations: a
                  bridge between dark energy and dark matter
                  problems'', {\it astro-ph/0411114}.
\bibitem{noodb}Nojiri, S \& Odintsov, S D ``Modified gravity with ${\ln}
               R$ terms and cosmic acceleration'', {\it Gen. Rel. Grav.}\ {\bf 36}\
               (2004), 1765-1780.
\bibitem{nooda}Nojiri, S \& Odintsov, S D ``Modified gravity with
               negative and positive powers of the curvature: unification of the
               inflation and of the cosmic acceleration'', {\it Phys. Rev.}\ {\bf D68}\
               (2003), 123512.
\bibitem{chiba03}Chiba, T ``$1\over R$ gravity and scalar--tensor
                 gravity'', {\it Phys. Lett.}\ {\bf B575}\ (2003), 1-3.
\bibitem{sousawodard04}Soussa, M E \& Woodard, R P ``The force of
                       gravity from a Lagrangian containing inverse
                       powers of the Ricci scalar'', {\it Gen. Rel. Grav.}\
                       {\bf 36}\ (2004), 855-862.
\bibitem{dolgovkawasaki2003}Dolgov, A D \& Kawasaki, M ``Can modified gravity
                            explain accelerated cosmic expansion?'', {\it Phys. Lett.}\
                            {\bf B573}\ (2003), 1-4.
\bibitem{noodc}Nojiri, S \& Odintsov, S D ``The minimal curvature of
               the universe in modified gravity and conformal anomaly resolution
               of the instabilities'', {\it Mod. Phys. Lett.}\ {\bf A19}\
               (2004), 627-638.
\bibitem{maba}Madsen, M S \& Barrow, J D ``De Sitter ground states
              and boundary terms in generalized gravity'', {\it \NP}\
              {\bf B323}\ (1989), 242-252.
\bibitem{marza}Mardones, A \& Zanelli, J ``Lovelock--Cartan theory
               of gravity'', {\it \CQG}\ {\bf 8}\ (1991),
               1545-1558.
\bibitem{palastepe}Palatini, A  ''Deduzione invariantiva delle
                  equazioni gravitazionali dal principio di Hamilton'', {\it Rend. Circ.
                  Mat. Palermo}\ {\bf 43}\ (1919), 203-212;\newline
                  Stephenson, G ``Variational principles \& gauge theories of
                  gravitation'', {\it \JP}\ {\bf A10}\ (1977), 181-184.
\bibitem{buchde}Buchdahl, H A ``Quadratic Lagrangians and Palatini's
                device'', {\it \JP}\ {\bf A12}\ (1979), 1229-1234.
\bibitem{bshs}Shahid-Saless, B ''Palatini variation of curvature
              squared action and gravitational collapse'', {\it
              \JMP}\ {\bf 32}\ (1991), 694-697.
\bibitem{mengwangsotiriou}Meng, X-H, \& Wang, P ``$R^2$ corrections to the
                          cosmological dynamics of inflation in the
                          Palatini formulation'', {\it \CQG}\ {\bf 21}\
                          (2004), 2029-2036; ``Palatini formulation of the
                          $R^{-1}$ modified gravity with an additionally
                          squared scalar curvature term'', {\it ibid.}\
                          {\bf 22}\ (2005), 23-32;\newline
                          Sotiriou, T P ``Unification of inflation and cosmic
                          acceleration in the Palatini formalism'', {\it
                          gr-qc/0509029}.
\bibitem{volikmengwangcapoetal}Vollick, D N ``$1\over R$
                               curvature corrections as the source
                               of the cosmological acceleration'', {\it Phys. Rev.}\
                               {\bf D68}\ (2003), 063510;\newline
                               Meng, X-H, \& Wang, P ``Modified
                               Friedmann equations in
                               $R^{-1}$--modified gravity'', {\it \CQG}\ {\bf 20}\
                               (2003), 4949-4962; ``Cosmological
                               evolution in $1\over R$ gravity
                               theory'', {\it ibid.}\ {\bf 21}\
                               (2004), 951-960;\newline
                               Capozziello, S, Cardone, V F \& Francaviglia, M
                               ``$f(R)$ theories of gravity in Palatini matched
                               with observations'', {\it astro-ph/0410135}.
\bibitem{olmok04}Olmo, G J \& Komp, W ``Non--linear gravity theories
                 in the metric and Palatini formalisms'', {\it
                 gr-qc/0403092}.
\bibitem{flanaganmengwang}Flanagan, E E ``Palatini form of $1\over
                          R$ gravity'', {\it \PRL}\ {\bf 92}\ (2004),
                          071101; ``Higher order gravity theories and scalar
                          tensor theories'', {\it \CQG}\ {\bf 21}\ (2004),
                          417-426;\newline
                          Meng, X-H, \& Wang, P ``Palatini formulation of modified
                          gravity wih squared scalar curvature'', {\it
                          astro-ph/0308284}; ``Palatini formulation of modified
                          gravity with ${\ln} R$ terms'', {\it Phys. Lett.}\
                          {\bf B584}\ (2004), 1-7.
\bibitem{vollick2004}Vollick, D N ``On the viability of the
                     Palatini form of $1\over R$ gravity'', {\it \CQG}\ {\bf 21}\
                     (2004), 3813-3816.
\bibitem{kermadhost}Kerner, R ``Cosmology without singularity and
                    nonlinear gravitational Lagrangians'', {\it \GRG}\
                    {\bf 14}\ (1982), 453-469;\newline
                    Madore, J ``On the nature of the initial singularity in
                    a Lanczos cosmological model'', {\it \PL}\ {\bf 111A}\
                    (1985), 283-284;\newline
                    Horowitz, G T \& Steif, A R ``Spacetime singularities in
                    string theory'', {\it \PRL}\ {\bf 64}\ (1990), 260-263.
\bibitem{baotmuss}Barrow, J D \& Ottewill, A C ``The stability of
                  general relativistic cosmological theory'', {\it \JP}\
                  {\bf A16}\ (1983), 2757-2776;\newline
                  M\"uller, V, Schmidt, H-J \& Starobinsky A A ``The stability
                  of the de Sitter space--time in fourth order
                  gravity'', {\it \PL}\ {\bf B202}\ (1988),
                  198-200.
\bibitem{faraoni05}Faraoni, V ``Modified gravity and the stability
                   of de Sitter space'', {\it \PR}\ {\bf D72}\
                   (2005), 061501.
\bibitem{tey}Teyssandier, P ``Linearized $R+R^2$ gravity: a new
             gauge and new solutions'', {\it \CQG}\ {\bf 6}\ (1989),
             219-229.
\bibitem{qus}Quant, I \& Schmidt, H-J ``The Newtonian limit of
             fourth and higher order gravity'', {\it Astron. Nachr.}\
             {\bf 312}\ (1991), 97-102.
\bibitem{dmwdbco3afrts}Dick, R ``On the Newtonian limit in gravity models with
                       inverse power of $R$'', {\it \GRG}\ {\bf 36}\
                       (2004), 217-224;\newline
                       Meng, X-H, \& Wang, P ``Gravitational potential in
                       Palatini formulation of modified gravity'', {\it Gen.
                       Rel. Grav.}\ {\bf 36}\ (2004), 1947-1954;\newline
                       Dom\'inguez, A E \& Barraco, D E
                       ``Newtonian limit of the singular $f(R)$
                       gravity in the Palatini formalism'', {\it \PR}\ {\bf D70}\
                       (2004), 043505;\newline
                       Capozziello, S ``Newtonian limit of extended
                       theories of gravity'', {\it gr-qc/0412088};\newline
                       Olmo, G J ``The gravity Lagrangian according to solar
                       system experiments'', {\it gr-qc/0505101}; ``Post--Newtonian
                       constraints on $f(R)$ cosmologies in metric formalism'',
                       {\it gr-qc/0505135}; ``Post--Newtonian constraints on $f(R)$
                       cosmologies in Palatini formalism'', {\it gr-qc/0505136};\newline
                       Allemandi, G, Francaviglia, M, Ruggiero, M
                       L \& Tartaglia, A ``Post--Newtonian parameters from alternative
                       theories of gravity'', {\it gr-qc/0506123};\newline
                       Sotiriou, T P ``The nearly Newtonian regime in non--linear
                       theories of gravity'', {\it gr-qc/0507027}.
\bibitem{cembra05}Cembranos, J A R ``The Newtonian limit at
                  intermediate energies'', {\it gr-qc/0507039}.
\bibitem{capoztroi05}Capozziello, S \& Troisi A ``PPN--limit of
                     fourth order gravity inspired by
                     scalar--tensor gravity'', {\it gr-qc/0507545}.
\bibitem{teto}Teyssandier, P \& Tourrenc, Ph ``The Cauchy problem
              for the $R+R^2$ theories of gravity without torsion'',
              {\it \JMP}\ {\bf 24}\ (1983), 2793-2799.
\bibitem{strmaluab}Strominger, A ''Positive energy theorem for $R+R^2$
                   gravity'', {\it \PR}\ {\bf D30}\ (1984), 2257-2259;\newline
                   Maluf, W ``Energy spectrum of quadratic theories
                   of gravitation'', {\it \CQG}\ {\bf 6}\ (1989),
                   1189-1195; ``Positivity of energy of $R+R^2$ theories of
                   gravitation'', {\it ibid.}, L151-L154.
\bibitem{magsok}Magnano, G \& Sokolowski, L M ''On physical
                equivalence between non-linear gravity theories
                and a general relativistic self-gravitating scalar
                field'', {\it \PR}\ {\bf D50}\ (1994), 5039-5059.
\bibitem{mffabsirz}Magnano, G, Ferraris, M \& Francaviglia, M
                   ``Non-linear gravitational Lagrangians'', {\it \GRG}\
                   {\bf 19}\ (1987), 465-479;
                   ``Legendre transformation and dynamical structure of
                   higher derivative gravity'', {\it \CQG}\ {\bf 7}\
                   (1990), 557-570;\newline
                   Sirousse Zia, H ``Singularity theorems and the
                   [general relativity + additional matter fields]
                   formulation of metric theories of gravitation'',
                   {\it \GRG}\ {\bf 26}\ (1994), 587-597.
\bibitem{bran}Brans, C H ``Non-linear Lagrangians and the significance of
              the metric'', {\it \CQG}\ {\bf 5}\ (1988), L197-L199.
\bibitem{ffmabso}Ferraris, M, Francaviglia, M \& Magnano, G ``Do non-linear
                 metric theories of gravitation really exist?'', {\it \CQG}\
                 {\bf 5}\ (1988), L95-L99;
                 ``Remarks on the physical metric in non-linear theories of
                 gravitation'', {\it ibid.}\ {\bf 7}\ (1990), 261-263;\newline
                 Sokolowski, L M ''Physical versions of non-linear gravity theories
                 and positivity of energy'', {\it \CQG}\ {\bf 6}\ (1989), 2045-2050.
\bibitem{pasa}Pascual-S\'anchez, J-F ``Variational principles and quantum
              gravity'' in {\it Recent Developments in Gravitation}, Proc.
              {\it Relativity Meeting}, Barcelona, Spain, Sept. 1989, Eds.
              E Verdaguer, J Garriga \& J C\'espedes (World Scientific,
              Singapore, 1990), pp 397-404.
\bibitem{bacomaewandhiow}Barrow, J D \& Cotsakis, S ''Inflation and the conformal
                         structure of higher order gravity theories'', {\it \PL}\
                         {\bf B214}\ (1988), 515-518;\newline
                         Maeda, K ``Towards the Einstein--Hilbert action via
                         conformal transformation'', {\it \PR}\ {\bf D39}\
                         (1989), 3159-3162;\newline
                         Wands, D ``Extended gravity theories and the
                         Einstein--Hilbert action'', {\it \CQG}\ {\bf 11}\ (1994),
                         269-280;\newline
                         Hindawi, A, Ovrut, B A \& Waldram, D ``Non-trivial vacua in
                         higher derivative gravitation'', {\it \PR}\ {\bf D53}\
                         (1996), 5597-5608.
\bibitem{goss}Gottl\"ober, S, Schmidt, H-J \& Starobinsky, A A
              ``Sixth-order gravity and conformal transformations'',
              {\it \CQG}\ {\bf 7}\ (1990), 893-900.
\bibitem{bdcsnoocnocacn}Boulware, D G \& Deser, S ``String--generated gravity
                       models'', {\it \PRL}\ {\bf 55}\ (1985), 2656;\newline
                       Cai, R G \& Soh, K S ``Topological black holes in the
                       dimensionally continued gravity'', {\it \PR}\
                       {\bf D59}\ (1999), 044013;\newline
                       Nojiri, S, Odintsov, S D \& Ogushi, S
                       ``Cosmological and black hole brane--world
                       universe in higher derivative gravity'', {\it \PR}\
                       {\bf D65}\ (2002), 023521;\newline
                       Cvetic, M, Nojiri, S \& Odintsov, S D
                       ``Black hole thermodynamics and negative
                       entropy in de Sitter and anti--de Sitter
                       Einstein--Gauss--Bonnet gravity'', {\it \NP}\
                       {\bf B628}\ (2002), 295-330;\newline
                       Cai, R G ``Gauss--Bonnet black holes in AdS spaces'',
                       {\it \PR}\ {\bf D65}\ (2002), 084014;\newline
                       Cho, Y M \& Neupane, I P ``Anti--de Sitter black holes,
                       thermal phase transition and holography in higher
                       curvature gravity'', {\it \PR}\ {\bf D66}\ (2002),
                       024044.
\bibitem{mtbnovrett}Maartens, R \& Taylor, D R ``Fluid dynamics in higher order
                    gravity'', {\it \GRG}\ {\bf 26}\ (1994),
                    599-613;\newline
                    Rippl, S, van Elst, H, Tavakol, R \& Taylor, D ``Kinematics
                    and dynamics of $f(R)$ theories of gravity'', {\it \GRG}\
                    {\bf 28}\ (1996), 193-205;\newline
                    Brevik, I, Nojiri, S, Odintsov, S D \& Vanzo, L
                    ``Entropy and universality of Cardy--Verlinde
                    formula in dark energy universe'', {\it \PR}\ {\bf D70}\
                    (2004), 043520.
\bibitem{buchd}Buchdahl, H A ``\"Uber die Variationsableitung von
               fundamental-invarianten beliebig hoher Ordnung'',
               {\it  Acta Math.}\ {\bf 85}\ (1951), 63-72.
\bibitem{aetal}Amendola, L, Mayer, A B, Capozziello, S, Gottl\"ober, S,
               M\"uller, V, Occhionero, F \& Schmidt, H-J ``Generalized
               sixth order gravity and inflation'', {\it \CQG}\
               {\bf 10}\ (1993), L43-L47.
\bibitem{stepd}Stephenson, G ``Variational principles for the
               gravitational field'', {\it Lett. \NC}\ {\bf 1}\ (1969),
               97-99.
\bibitem{roxcol}Roxburgh, I W ``Non-linear Lagrangian theories of gravity'',
                {\it \GRG}\ {\bf 8}\ (1977), 219-225;\newline
                Coley, A A ``Homothetic vectors and higher order Lagrangian
                theories of gravity'', {\it \CQG}\ {\bf 6}\ (1989),
                1213-1218.
\bibitem{farb}Farhoudi, M ``Lovelock tensor as generalized Einstein
              tensor'', {\it gr-qc/9510060}.
\bibitem{farc}Farhoudi, M ``Classical trace anomaly'', {\it Int. J. Mod.
              Phys. D}\ {\bf 14}\ (2005), 1233-1250.
\bibitem{dufc}Duff, M J ``Twenty years of the Weyl anomaly'', {\it \CQG}\
              {\bf 11}\ (1994), 1387-1403.
\bibitem{asgosh}Asorey, M, Gorbar, E V \& Shapiro, I L
                ``Universality and ambiguities of the conformal
                anomaly'', {\it \CQG}\ {\bf 21}\ (2003), 163-178.
\bibitem{dufa}Duff, M J ``Observations on conformal anomalies'',
              {\it \NP}\ {\bf B125}\ (1977), 334-348.
\bibitem{bocr}Bonora, L, Cotta-Ramusino, P \& Reina, C ``Conformal
              anomaly and cohomology'', {\it \PL}\ {\bf 126B}\ (1983),
              305-308.
\bibitem{bopb}Bonora, L, Pasti, P \& Bregola, M ``Weyl cocycles'',
              {\it \CQG}\ {\bf 3}\ (1986), 635-649.
\bibitem{noodog}Nojiri, S, Odintsov, S D \& Ogushi, S
                ``Holographic renormalization group and conformal
                anomaly for AdS$_9$/CFT$_8$\ correspondence'', {\it \PL}\
                {\bf B500} (2001), 199-208.
\bibitem{ddichr}Deser, S, Duff, M J \& Isham, C J ``Non-local conformal
                anomalies'', {\it \NP}\ {\bf B111}\ (1976),
                45-55;\newline
                Christensen, S M ``Regularization, renormalization, and
                covariant geodesic point separation'', {\it \PR}\
                {\bf D17}\ (1978), 946-963.
\bibitem{avra}Avramidi, I G ``New algebraic methods for calculating the
              heat kernel and effective action in quantum gravity and
              gauge theories'' in {\it Heat Kernel Techniques and Quantum
              Gravity}, Discourses in Mathematics and its Applications, Ed. S A Fulling,
              (Department of Mathematics, Texas A\&M University, 1995), pp 115-140,
              {\it gr-qc/9408028}.
\bibitem{frw}Fulton, T, Rohrlich, F \& Witten, L ``Conformal invariance in
             physics'', {\it \RMP}\ {\bf 34}\ (1962), 442-457.
\bibitem{mulhc}M\"uller-Hoissen, F ``Spontaneous compactification with
               quadratic and cubic curvature term'', {\it \PL}\ {\bf 163B}\
               (1985), 106-110.
\bibitem{fard}Farhoudi, M {\it Non-linear Lagrangian Theories of Gravitation},
              (Ph.D. Thesis, Queen Mary \& Westfield College, University of
              London, 1995).
\bibitem{whelt}Wheeler, J T ``Symmetric solutions to the Gauss--Bonnet
               extended Einstein equations'', {\it \NP}\ {\bf B268}\ (1986),
               737-746.
\bibitem{farf}Farhoudi, M ``New derivation of Weyl invariants in six
              dimensions'', work in progress.
\bibitem{defa}Deruelle, N \& Fari\~na-Busto, L ``The Lovelock gravitational
              field equations in cosmology'', {\it \PR}\ {\bf D41}\ (1990), 3696-3708.
\bibitem{desch}Deser, S \& Schwimmer, A ``Geometric classification of
               conformal anomalies in arbitrary dimensions'', {\it \PL}\ {\bf
               B309}\ (1993), 279-284.
\bibitem{hwkvyon}Hamber, H W \& Williams, R M ``Two dimensional simplical quantum
                 gravity'', {\it \NP}\ {\bf B267}\ (1986), 482-496;\newline
                 Katanayev, M O \& Volovich, I V ``Two dimensional gravity
                 with dynamical torsion and string'', {\it \AP}\ {\bf 197}\ (1990),
                 1-32;\newline
                 Yoneya, T ``Higher derivative quantum gravity in two dimensions'',
                 {\it \PL}\ {\bf 149B}\ (1984), 111-116.
\bibitem{bro}Brown, J D, {\it Lower Dimensional Gravity}, (World Scientific,
             Singapore, 1988).
\bibitem{sta}Stachel, J ``Einstein's struggle with general covariance,
             1912-1915'' presented at {\it General Relativity and
             Gravitation}~9$^{th}$, 1980 at Jena, Germany; reprinted as ``Einstein's
             search for general covariance, 1912-1915'' in Proc. {\it Einstein and
             the History of General Relativity}, Osgood Hill Conference,
             Massachusetts, May 1986, Eds. D Howard \& J Stachel (The Center for
             Einstein Studies, Boston Univ., 1989), pp 63-100.
\bibitem{einleinll}Einstein, A, {\it Wrote to:} P Ehrenfest, on $26^{th}$ December,
                   1915, EA 9-363;\newline
                   {\it Wrote to:} M Besso, on $3^{rd}$ January, 1916 in {\it
                   Albert Einstein, Michele Besso, Correspondence, 1903-1955},
                   Ed. P Speziali (Hermann, Paris, 1972), pp 63-64.
\bibitem{eindd}Einstein, A ``Relativity and the problem of space'' (1952),
               Appendix~5 in {\it Relativity, the Special and the General Theory:
               A Popular Exposition}, translated by: R W Lawson (Methuen, London,
               $15^{th}$ edition 1954), pp 135-157.
\bibitem{adl}Adler, I, {\it A New Look at Geometry}, (John Day
             Com., New York, 1966).
\bibitem{mashliuwess94}Mashhoon, B, Liu, H \& Wesson, P S
                       ``Space--time--matter'' in Proc.~7$^{th}$
                       {\it Marcel Grossmann Meeting}, Stanford, 1994,
                       pp 333-335.
\bibitem{gra}Graves, J C, {\it The Conceptual Foundation of Contemporary
             Relativity Theory}, (MIT Press, 1971).
\bibitem{rysh}Ryan, Jr., M P \& Shepley, L C, {\it Homogeneous
              Relativistic Cosmologies}, (Princeton Univ. Press, 1975).
\bibitem{schntu}Schwarzchild, k ``\"Uber das gravitationsfeld
                eines massenpunktes nach der Einsteinschen
                theorie'', {\it Sitzber. Preuss. Akad. Wiss.
                Berlin}\ (1916), pp 189-196;\newline
                Taub, A H ``Empty space--times admitting a three
                parameter group of motions'', {\it \AM}\ {\bf 53}\
                (1951), 472-490;\newline
                Newman, E T, Tamburino, L \& Unti, T J
                ``Empty--space generalization of the Schwarzchild
                metric'', {\it \JMP}\ {\bf 4}\ (1963), 915-923.
\bibitem{abbasi}Abbassi, A M {\it Revisiting the Mach's and Correspondence
                Principle in General Relativity and Concept of Inertia},
                (Ph.D. Thesis, Tarbiat Modarres University, Tehran,
                Iran, 2001).
\bibitem{sal}Salam, A ``Gauge unification of fundamental forces'', {\it
             \RMP}\ {\bf 52}\ (1980), 525-538.
\bibitem{rtz}Romero, C, Tavakol, R \& Zalaletdinov, R ``The
             embedding of general relativity in five dimensions'',
             {\it \GRG}\ {\bf 28}\ (1996), 365-376.
\bibitem{wesson99and2005}Wesson, P S, {\it Space--Time--Matter, Modern
                         Kaluza--Klein Theory}, (World Scientific, Singapore,
                         1999);\newline
                         Wesson, P S ``In defence of Campbell's
                         theorem as a frame for new physics'',
                         {\it gr-qc/0507107}.
\bibitem{farh}Farhoudi, M, work in progress.
\end{thebibliography}
\end{document}